\documentclass[conference, 10pt]{IEEEtran}
\usepackage{amssymb,amsmath}
\usepackage{setspace}
\usepackage{graphicx}
\usepackage{epsfig,cite,latexsym,subfigure}
\usepackage{theorem}
\usepackage{times}
\usepackage{epsf,cite,subfigure}
\usepackage{amsmath,amssymb}

\usepackage{floatflt}
\usepackage{url}
\usepackage{times}
\usepackage[usenames]{color}
\usepackage{dsfont}
\usepackage[final]{pdfpages}

\newcommand{\set}[1]{\left[#1\right]}
\newcommand     {\paren}[1]{\left(#1\right)}

\title{{\Large AutoEncoders for Training Compact Deep Learning RF Classifiers for Wireless Protocols}}\vspace{-2mm}
\author{
\IEEEauthorblockN{\normalsize Silvija Kokalj-Filipovic, Rob Miller, Joshua Morman} \\\vspace{-2mm}\\
\IEEEauthorblockA{\small Perspecta Labs, Inc \\
\small\em \{skfilipovic, rmiller, jmorman\}@perspectalabs.com}}%, jsamson@ltsnet.net}} 
\begin{document}
\maketitle
\begin{abstract}
We show that \emph{compact} fully connected (FC) { deep learning} networks trained to classify wireless protocols using a hierarchy of { multiple} denoising autoencoders (AEs) outperform reference FC networks trained in a typical way, i.e., with a stochastic gradient based optimization of a given FC architecture. 
Not only is the complexity of such FC network, measured in number of trainable parameters and scalar multiplications, much lower than the reference FC and residual models, its accuracy also outperforms both models for nearly all tested SNR values (0 dB to 50dB). 
Such AE-trained networks are suited for { in-situ} protocol inference performed by simple mobile devices based on noisy signal measurements.  Training is based on the data transmitted by real devices, and collected in a controlled environment, and systematically augmented by a policy-based data synthesis process by adding to the signal any subset of impairments commonly seen in a wireless receiver.%Further research will extend this study to apply non-white noise in DAEs and to test the robustness and information loss of other signal transforms.
\end{abstract}
\vspace{-2mm}
\section{Intro}\vspace{-1mm}
A new research direction is emerging in the field of wireless communications, aiming to emulate and possibly replace certain signal processing algorithms by deep learning (DL) models. We here present a system based on DL that can train neural network \emph{(NN)} models to perform \emph{Industrial, Scientific and Medical (ISM) band} wireless technology (protocol) classification using 2.4 GHz ISM wideband spectrum samples. The samples are generated by four ISM emitter classes: WiFi 802.11n \cite{80211n}, Bluetooth (BT) \cite{BT}, ZigBee \cite{ZBee} and NRF \cite{NRF}. 
 The contributions of this work are a novel way to collect, augment and curate radio frequency  \emph{(RF)} data, as well as a DL training framework based on the {\emph stacked denoising autoencoder} \emph{(SDAE)}, inspired by the results in \cite{VincentSDAE} that established the value of using SDAE to learn  higher level representations. 
The system for data curation is based on low-noise (clean) wireless signals, carefully recorded  in a controlled environment (Fig.~\ref{fig:f0}), and allows for gradual and tractable distortion of such data according to Fig.~\ref{fig:f1}. The figure depicts the data synthesis system which builds the distorted (augmented) dataset from the clean samples stored in a SigMF database \cite{SigMF}.  
While we tested NN models of various complexity on this classification task, the model proposed in this paper is { a compact}  FC model of small complexity amenable to deployment in mobile devices with constrained resources. To achieve high accuracy across a wide range of signal SNRs with a compact FC model we trained the model using an SDAE.  Finally, it is important to emphasize that datapoints used for DL training are very short bursts of RF signal samples, about $1 \mu$s long.%, which will allow for real time in-situ classification by small field devices that combine spectrum sensing and the proposed DL classifier.
\subsubsection{Existing Work}
The research in the area of RF-based DL of the PHY layer is still embryonic \cite{PHYDeep}. Modulation recognition \emph{(ModRec)} is the most popular application of DL here. Most of the existing work is based on convolutional (CNN) architectures \cite{ConvDeepRF}.  Paper \cite{OTADeepRF} features an in-depth study on the performance of DL ModRec methods on Over-the-Air (OTA) captured RF communication signals synthetically designed in Software Defined Radio (SDR). Note that we use RF samples from real commercial devices. The paper \cite{OTADeepRF} demonstrates that in the ModRec context DL provides significant performance benefits compared to conventional feature extraction methods. Apart from exploring optimal DL architectures and comparing their classification accuracy with state-of-the-art performance based on signal cumulants or their cyclo-stationary properties \cite{cyclostat}, ModRec research introduced the problem of collecting and pre-processing RF datasets for training and evaluating deep classifiers, e.g., \cite{GNUdataset}. The majority of the existing ModRec work trains DL networks utilizing complex baseband samples at a sampling rate close to Nyquist, much lower than we use here. %Another common application of DL on RF data is specific emitter identification. The paper \cite{Merchant} by Merchant, et al., and the references therein,  explore the use of DL based on OTA data to detect PHY-layer attributes for the identification of radio devices. 
\subsubsection{Departing from Modulation Recognition}
Our classification task, which we refer to as {\em ProtRec},  is different from ModRec. Real wireless protocols have many more distinguishing features, allowing 2 different protocols to utilize the same modulation (e.g., BT and NRF both use GFSK). %They use different packet framing, synchronization methods, rules for in-band signaling, discernible resource allocation in time and frequency, to mention just a few differences. 
Macro features, such as frequency hopping and packet framing, although distinguishing, are not utilized in ProtRec classification as datapoints would have to cover much longer periods of time to capture them. We used short burst of RF samples that cover a couple of symbols.  { Although very short, they }may include both the modulation used in the preamble and modulation specific to the payload. { Also, despite the short duration, there is} an imbalance in the number of symbols included in these 128 sample long bursts: while a datapoint is roughly equivalent to one symbol with BT, with WiFi it spans many symbols. { Since we want our classifier to perform accurately regardless of the data carried by the RF signals we utilized different traffic profiles during datataset collection. Hence, dataset statistics may vary depending on the traffic carried by the packet payload.  We estimated that the diversity of traffic carried by the payloads has been integrated out due to the number of bursts per class ($>$ 20K) and random sampling.} These issues, and the fact  that the data is emitted by real devices makes perfect classification difficult. This paper describes the approach we took to train compact protocol classifiers  based on OTA data. Organizationally, the next section describes data collection and preprocessing in preparation for DL, followed by considerations for selecting DL models for protocol classification, including the role of autoencoders (AEs). { Section~\ref{sec:results} presents the procedure and relevant results such as accuracy over training epochs to illustrate convergence of the proposed model, inference accuracy for different SNR levels to illustrate robustness, and the confusion matrices for extreme SNRs to illustrate what is the most difficult to learn.} Note that we did not include loss curves due to space considerations, but  for all featured NNs the training converged w/o overfitting.%\vspace{-2mm} %We conclude in the last section.
\vspace{-2mm}
\section{Problem Definition and System Description}\vspace{-1mm}
Our goal is to classify wideband RF samples from the 2.4 GHz ISM band into 4 protocol classes using a compact DL classifier. The data is obtained using an SDR receiver to capture RF transmissions of several commercial ISM modems. %The training is done on computationally strong devices based on GPUs, but the inference is to be performed on weak mobile devices. Hence, our training can be complex, but the resulting DL network should be compact. { Compactness of the model here primarily means small number of scalar multiplications per inference.}
We next describe the issues related to data collection and synthesis, and NN architecture selection and training. 
%{Information-theoretically speaking, they argued
%that the DNN learning problem can be posed as extracting
%the minimal sufficient statistics of input data with respect to
%the output, which can be viewed as a special case in the rate
%distortion theory: the information bottleneck (IB) method [3]. 
%Since each layer depends only on the previous layer, the
%layers in a DNN form a Markov chain Y−X−X1−· · ·−XL.
%According to the data processing inequality, the following
%holds for any $j \geq i$:
%$I(Y; X)  \geq I(Y; X_i
%)  \geq I(Y; X_j
%)  \geq I(Y; \hat{Y} ),$ 
%where I(; ) is the Shannon mutual information. The equality
%hold if and only if each layer is a sufficient statistics of its
%input.
%} 
\subsubsection{Data Collection, Curation  and Preprocessing}\label{subsubsec:data}
\begin{figure}[t] %FIGURE 0
\begin{center}
\hspace{-5mm} \includegraphics [width=2.4in]{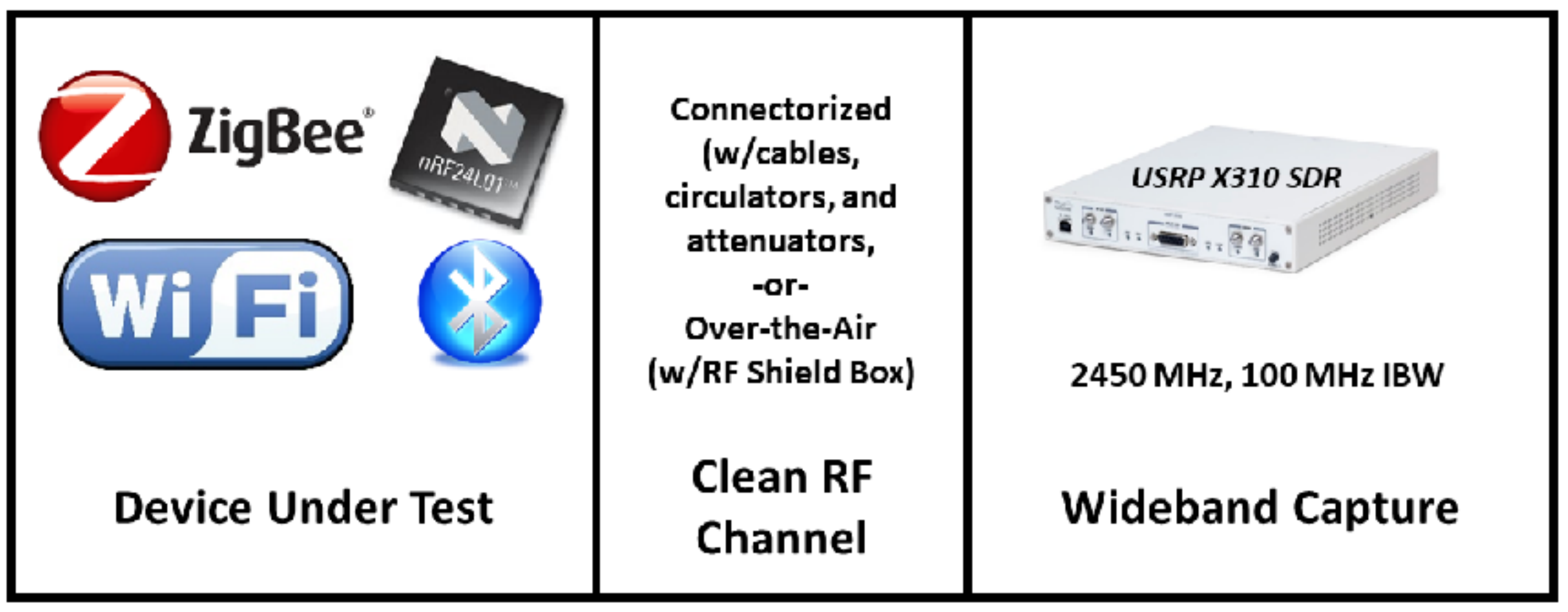}\vspace{-2mm}%\\
%\hspace{-5mm} \includegraphics [width=2.4in]{sampling.eps}\vspace{-2mm} %sampling.eps
%\caption{{ Collection of clean wideband RF signal samples - connectorized collection shown in the bottom}}\vspace{-4mm}   \label{fig:f0}
\caption{{ Collection of clean wideband RF signal samples}}\vspace{-4mm}   \label{fig:f0}
\end{center}
\end{figure}

We collected data from 5 devices per protocol, 4 from a common manufacturer, the fifth from a different one. %We utilized connectorized SDR receiver(s) with sample rate of $100MHz$ and center frequency of $2450 MHz$. Data collection is performed through a circulator while the transmitter was communicating with a peer device normally (OTA), e.g., transmitting a file or repeatedly pinging a receiver.  Please see Fig.~\ref{fig:f0}. 
Where possible, we used connectorized devices in conjunction with cables, circulators and attenuators to ensure pristine RF data.  For devices that necessitated OTA operation, we used an RF shield box to simplify the RF channel and minimize co-channel interference.  The SDR was a USRP X310 w/UBX RF daughterboard, with a sample rate of 100 MHz centered at 2450 MHz (the middle of the 2.4 GHz ISM band). Please see Fig.~\ref{fig:f0}.  
As opposed to a common approach to data collection for DL, we are not sampling the signals in the packet preamble only. Instead, the collection time covers several full packets, and then bursts of I/Q samples are extracted from the collected time-series of samples at random points. The training and evaluation datasets contain a large number of such bursts to make sure that traffic-related diversity is captured.
 
The small noise in the collected data is estimated using traditional signal processing, which confirmed that collected samples are of very high SNR, and can be considered clean. 
To facilitate data curation in terms of storage and manipulation, some basic pre-processing is needed. We detect and extract emissions, discarding the noise samples in between. With such temporal reduction we can store and process more information. We use a temporal buffer to guarantee capture of transitional regions, right before the rising edge of the emission. 

\begin{figure}[t] %FIGURE 1
\begin{center}
\hspace{-5mm} \includegraphics [width=3.0in]{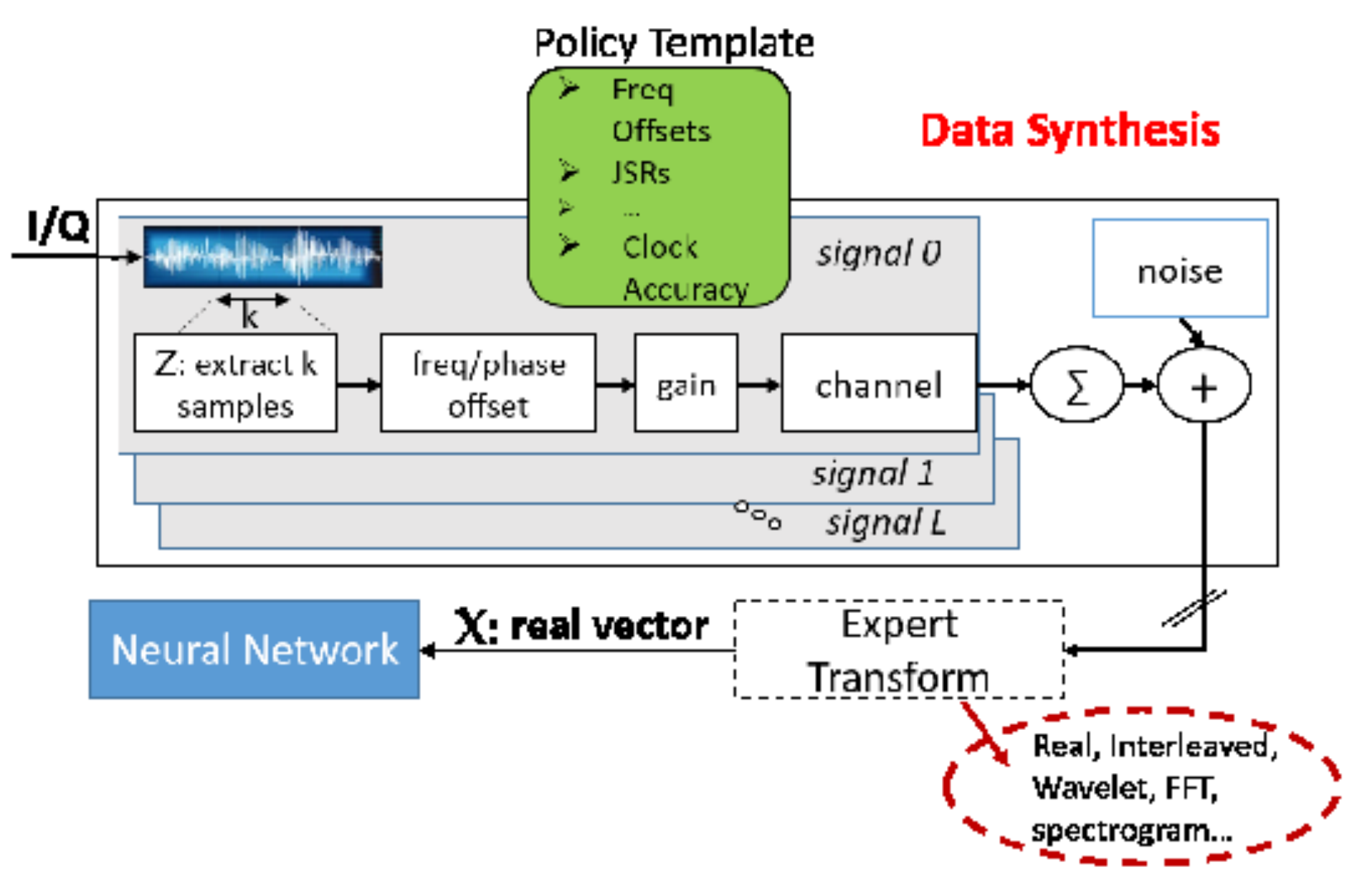}\vspace{-2mm}
\caption{Data synthesis system builds the distorted and transformed dataset from the clean samples} \vspace{-6mm}  \label{fig:f1} %\vspace{-5mm}
\end{center}
\end{figure}

The next step was to use a carefully designed parameterizable synthesis process to add channel distortions and receiver imperfections due to receiver's synchronization, noise, dynamic range, etc.
Our synthesis system (Fig.~\ref{fig:f1}) is policy based and allows adding to the signal any subset of impairments commonly seen in a wireless receiver, such as additive white-noise corruption at the
receiver due to physical device sensitivity, or non-impulsive delay spread due to propagation effects on multiple
paths.
%\item{Carrier frequency offset (CFO)}: carrier phase/ frequency offset due to disparate local oscillators and
%motion (Doppler).
%\item{Symbol rate offset (SRO)}: symbol clock offset due to disparate clock sources and motion.
%\item{Thermal Noise}: additive white-noise corruption at the
%receiver due to physical device sensitivity.
%\end{itemize}
The system also allows for gain adjustment{ , which is important for }interference emulation due to concurrent emissions. 
For this { classification task,} we add both time and frequency jitter to the clean signal. Also, the frequency channel of the signal  is { altered} to avoid identifying a signal solely based on the frequency channels utilized during data collection. This is important especially for cognitive networks where frequency bands are used
opportunistically and an exact fixed carrier frequency is rarely
known a priori. { We alter the channel by either randomizing it or by converting it to baseband, similar to \cite{Merchant}}. Note that the achieved accuracy on non-frequency randomized datasets was higher because it utilized frequency as a feature, which with a-priori known protocol frequencies works as a discriminator, but our goal is transferability of the results rather than showcasing perfect classification accuracy. Also, by eliminating frequency channels and applying jitter, we remove a feature that
could easily be spoofed by a malicious transmitter. { In our research we applied both frequency channel elimination methods,  and compared their effects on deep learning, which will be showcased in this paper too.} Delay spread was not applied to this dataset as part of our systematic approach to building classifiers. It will be in the future work. Thermal noise was applied to both training and evaluation datasets to meet a certain SNR requirement, either to test the robustness of the trained classifier, or during its training.
Finally, we may apply to the preprocessed data a data transform, such as FFT, or Wavelet Transform, and then convert the transformed data to real vectors, with or without information loss. In this paper, we used interleaved I and Q samples which is a simple transform from complex to real set, i.e., for a vector $c$ of $k$ I/Q samples $c_1, \cdots, c_k$, where $c_i = x_i +j\ast y_i,$ the transformed vector has $2k$ real elements $\set{x_1, y_1, x_2, y_2, \cdots, x_k, y_k}.$  

%%%%%%%%%%%%%%%%%%%%%%%%%%%%%%%%%%%%%%%%%%%%%%%%%%%%%%%%%%%%%%
%%%%%%%%%%%%%%%%%%%%%%%%%%%%%%%%%%%%%%%%%%%%%%%%%%%%%%%%%%%%%%
%%%%%%%%%%%%%%%%%%%%%%%%%%%%%%%%%%%%%%%%%%%%%%%%%%%%%%%%%%%%%%
\subsubsection{Selecting DL Models for Protocol Classification}
Simply put, we have to address the following DL choices: 
%\begin{itemize}
$\left.1\right)$ what type of DL models to use (e.g., mainly composed of convolutional layers, or with FC layers, classical convolutional (CNN) models versus models composed of residual units (ResNets), the above mentioned models versus recurrent neural networks to leverage natural time correlation in the signal, etc); 
$\left.2\right)$ how complex the model should be, both in the number of parameters and multiplications;
$\left.3\right)$ how to combine supervised and unsupervised training to optimally extract salient features and obtain a robust classifier.
%\end{itemize}
These choices are all intertwined and depend on the complexity of the task: how many classes there are, and how separated datapoints of different classes are in the $k$-dimensional space. Here, dimensionality $k$ may be either the number of I and Q samples in a datapoint, or a smaller number, if we project the data into a lower-dimensional space, e.g., by using an AE. 

%It also depends on what our trade-offs are: if we prefer to trade the accuracy for a smaller footprint of the neural network, we will likely base the model on FC layers. FC models have smaller number of scalar multipliers for the same depth and width of the network.

{ In our prior work  we classified 16 different ISM-band signal classes, where some of the classes emerged from concurrent emissions. Hence, the classification complexity was much higher by both criteria, number of classes and separability. It called for models of higher capacity than FC networks, with ability to improve learning with depth. Theoretically, the accuracy would improve with increased depth. However, beyond a certain depth it drops due to vanishing gradients. ResNets \cite{ResNet} are deep architectures with direct identity mappings
between convolutional layers, which mitigate this phenomenon. %We found that ResNets outperform conventional CNNs in terms of computational efficiency, delivering high accuracy with additional layers but no overfitting or divergence. %The same conclusion was reached in \cite{OTADeepRF} for the modulation classification of OTA signals.
%Decreased information flow refers to the effect that repeated multiplication or convolution
%with randomly initialized weights has on the features from upper layers, making it harder for lower layers to learn salient information and gradient directions. 
%vanishing gradients and information loss. %allow for lossless propagation of forward information through the network. 
We opted to use them as a reference point for accuracy because we do not have to carefully optimize the number of layers, i.e., we can over-design the number of layers to guaranty maximum accuracy  without risking gradient instability \cite{KHeRes}.   

Our ResNets consist of residual units comprised of 2 1-dimensional convolutional layers and 1 identity layer, with batch normalization and dropouts applied to convolutional outputs. The number of layers and number of units per layer is parametrizable, as is the number of convolutional channels between layers, and the kernel size for each unit. 
Again, the ResNet design in itself is not the goal here except that it should be deep enough to have high learning capacity. The employed ResNet model, dubbed ISMResNet11, had 7 units per layer,  their kernel sizes from $3$ to $9,$ and total of 11 layers with 16 channels between convolutional layers. %, or 10 layers with kernel sizes from 5 to 33. The accuracy curve of ISMResNet11 is plotted as full line in Fig.~\ref{fig:f2}%~and~\ref{fig:f3}.
We will present different aspects of ISMResNet11 classification performance to establish the baseline for evaluation of the proposed compact model. 
For example, in Fig.~\ref{fig:f4a} we stacked confusion matrices obtained as we tested ISMResNet11 models trained on baseband transformed datapoints, one for random (top pane) and one for start bursts (bottom pane). Each confusion matrix is obtained using a dataset of  the indicated SNR level. Fig.~\ref{fig:f4a} shows that at lower SNRs WiFi performs best, while Bluetooth and NRF are being mixed up all the way to SNR of 40dB for random bursts and 30dB for start bursts. ZigBee becomes recognizable in between those SNR values. This is used as a reference to compare the accuracy per class obtained with the SDAE based classifiers trained on random/ start bursts with baseband channel transformation as illustrated in Fig.~\ref{fig:f4}
(prediction accuracy on 0 and 50dB signals, respectively). 
We also present the total accuracy of ISMResNet11 for all considered SNRs in Fig.~\ref{fig:f7}, which again shows the accuracy of all other models.

Apart from ISMResNet11, we needed a FC reference network for comparison with the FC network trained with SDAE. This is the regularly trained FC network of four layers, with 256, 128, 64, 32 (and 4 softmax) neurons  in its layers, respectively. It has 100000 multiplications (and trainable parameters), which roughly defines the  constraint for a small-footprint neural network that we seek.  %Figure\ref{fig:f5}%~and~\ref{fig:f6} 
% shows accuracy plot generated during training. 
The performance of the reference FC network was not satisfactory (below 80\% for high SNR using baseband-converted random bursts). We could probably add some more layers hoping to improve classification accuracy without overfitting. The alternative is what we propose in this paper: to train this network using AEs, expecting to be able to improve the accuracy by utilizing salient features extracted by the autoencoders (see subsection~\ref{subsubsec:auto}). This falls under the  $3^{rd}$ set of criteria influencing DL design. We can see in Fig.~\ref{fig:f4}~and ~\ref{fig:f7}  that by using the AE-based training we have not only been able to increase accuracy, but we also managed to lower the complexity of the FC network thus trained. Before addressing the AE-based training in Section~\ref{sec:results}, we next introduce basic types of AEs.}
%%%%%%%%%%%%%%%%%%%%%%%%%%%%%%%%%%%%%%%%%%%%%%%%%%%%%%
\begin{figure}[t] %FIGURE 1a
\begin{center}
\hspace{-5mm} \includegraphics [width=2.4in]{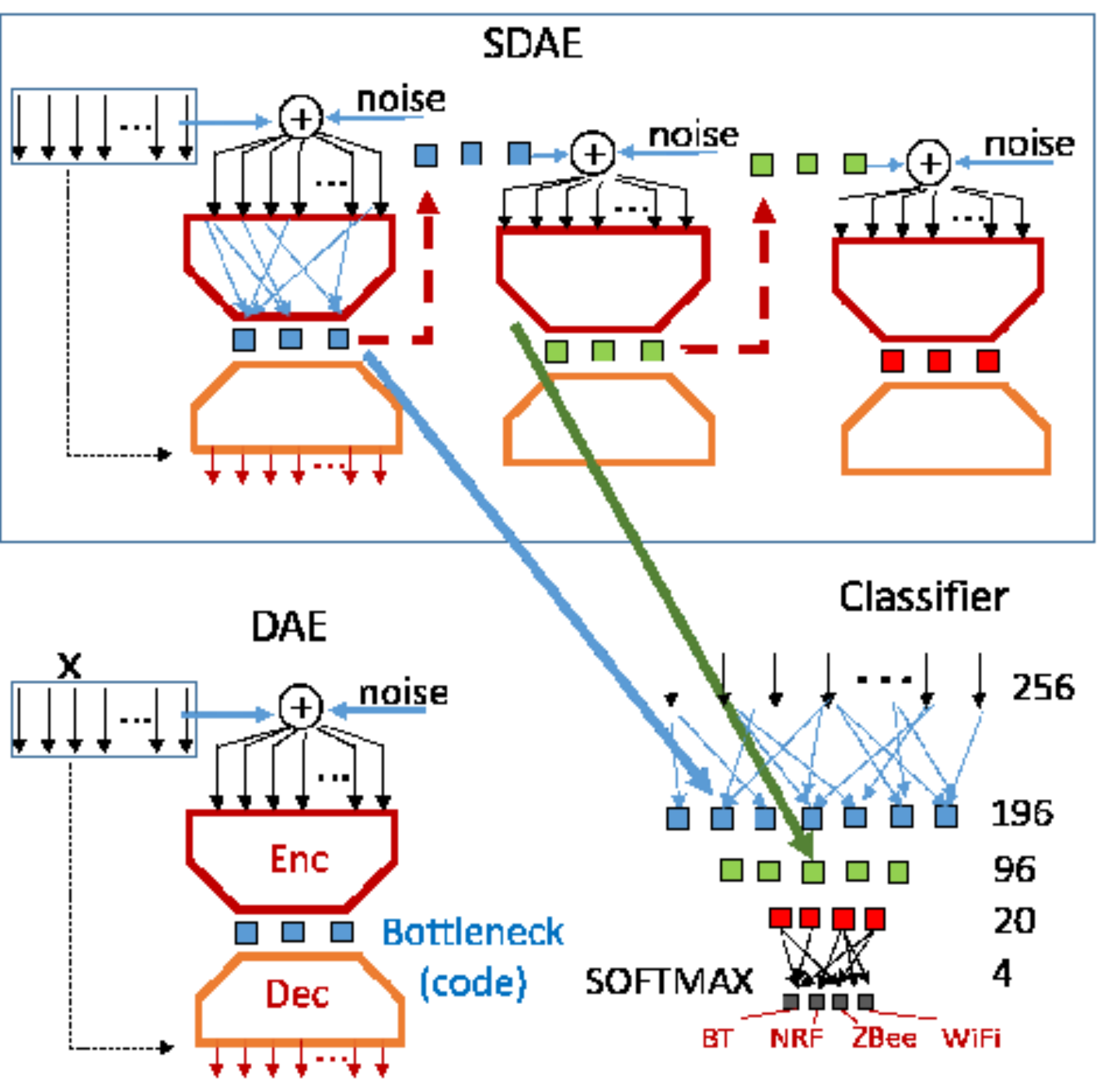}\vspace{-3mm}
\caption{SDAE is implemented by training three stacked AEs, each with independently corrupted inputs. The respective bottleneck layers are put together and refined within the classifier} \vspace{-6mm}  \label{fig:f1a}%\vspace{-6mm}
\end{center}
\end{figure}
%%%%%%%%%%%%%%%%%%%%%%%%%%%%%%%%%%%%%%%%%%%%%%%%%%%%%%
\subsubsection{Autoencoders}\label{subsubsec:auto}  
An autoencoder  (AE) is an unsupervised NN model which learns to extract data features leveraging an architecture that includes encoder and decoder. See Fig.~\ref{fig:f1a}, bottom left,  that shows the encoder-decoder structure of a denoising autoencoder (DAE), which is a special type of AE that corrupts input with noise, and forces the output to match the uncorrupted version of the input. Let us assume that the input is a vector of $k$ real  numbers $\set{x_1, \cdots, x_k}.$
The encoder maps data inputs to latent space $Z,$ and the AE is trained to allow the decoder to 
reconstruct the input from the latent space with  minimum reconstruction error $\mathcal{L}_r = \sum_{i=1}^{k}{\paren{x_i-o_i}^2}$, where the output of the AE is $O =\set{o_1, \cdots, o_k},$ and $o_i = \hat{x_i}.$  In case of DAE, the input is $\mathbf{y}=\set{x_1+n_1, \cdots, x_k+n_k}, $ and $n_k$ is an independent noise sample, usually  some Gaussian noise $n_k \sim \mathcal{N}\paren{\mu,\sigma}.$ The reconstruction loss function $\mathcal{L}_r$ is the same as with common AE.

The width of the bottleneck layer between encoder and decoder defines how compressed we want the data representation in the latent space to be, as we may use this bottleneck representation as an input to a classifier.
Extracting salient features through bottleneck compression leads to higher classification accuracy without prior expert knowledge. The AE bottleneck, trained into a  lower dimensional manifold of the dataset, also lowers the probability of overfitting the classifier. We accomplish both tasks using regularized stacked denoising autoencoders (SDAEs), as in Fig.~\ref{fig:f1a}. Similar use of SDAE in \cite{SparseDenoisingAEMigliori} to generalize training of a ModRec classifier  resulted in correct classification rates of 99\% at 7.5 dB SNR and 92\% at 0 dB SNR in a 6-way classification. Adding Gaussian noise to inputs randomly has an effect of forcing the network to not change the output for the input values in a ball around the exact input. This adds robustness against overfitting to the model. 
A complementary approach to avoid overfitting is to add a stochastic margin around the model weights. This could be achieved using dropouts (as is common in FC classifiers). However, we applied different weight regularization which enforces sparsity of bottleneck layers. The sparsity parameter $\rho$ defines the number of active neurons in the bottleneck layer, and we model it as a parameter of Bernoulli distribution, i.e., the probability of a neuron activation. The reconstruction loss $\mathcal{L}_r$ is regularized with a term equal to the Kullback-Leibler (KL) distance between $\rho$ and the actual mean activity of the bottleneck layer: $ \mathcal{L}=\mathcal{L}_r + \lambda KL\left(\rho || \hat{\rho}\right), $ where $$ KL\left(\rho|| \hat{\rho}\right) = \sum\limits_{j = 1}^{{k}} {\rho \log \frac{\rho }{{{{\hat \rho }_ j}}}}  + \left( {1 - \rho } \right)\log \frac{{1 - \rho }}{{1 - {{\hat \rho }_ j}}}.$$%%%%%%%%%%%%%%%%%%%%%%%%%%%%%%%%
%\begin{figure}[t] \vspace{-15mm}%FIGURE 1a 
%\begin{center}
%\begin{tabular}{c}
%\hspace{-5mm} \includegraphics [width=3.3in]{STAEClusterBegin131p0FRan9d.eps}\vspace{-15mm}\\%\vspace{-2mm}
%\hspace{-5mm} \includegraphics [width=3.3in]{STAEClusterOutNoise131p0FRan9d.eps}\vspace{-15mm}
%\end{tabular}
%\caption{3 strongest PCA-components of each bottleneck layer (one pane for each bottleneck) color-coded by the SDAE input (datapoint) class: Top graphs show this 3D represenation before DAE training and the bottom graphs are after the training is completed}\vspace{-4mm}   \label{fig:f1b}
%\end{center}
%\end{figure}
%%%%%%%%%%%%%%%%%%%%%%%%%%%%%%%%%%%%%%%%%%%%%%%%%%%%%%%%%%%%%%%%%%%%%%%%%%%%%%%%%%%%%%%%%%%%%%%%%%%%%%%%
The SDAE that we used to pre-train layers of the compact FC classifier  has 3 stages (Fig.~\ref{fig:f1a}). The $1^{st}$ uses the bursts of RF samples dusted with noise. Once it is trained, the values in the bottleneck layer (for each burst) are corrupted with independent noise, and fed into the next DAE, and so on. %\vspace{-2mm}%All three DAEs use independent Gaussian noise, and same regularization methods. 
%Details are addressed next.
\vspace{-2mm}\section{ISM Classification Procedure and Results}\label{sec:results}
\begin{figure}[t] %FIGURE 4a
\begin{center}
\hspace{-5mm} \includegraphics [width=3.0in]{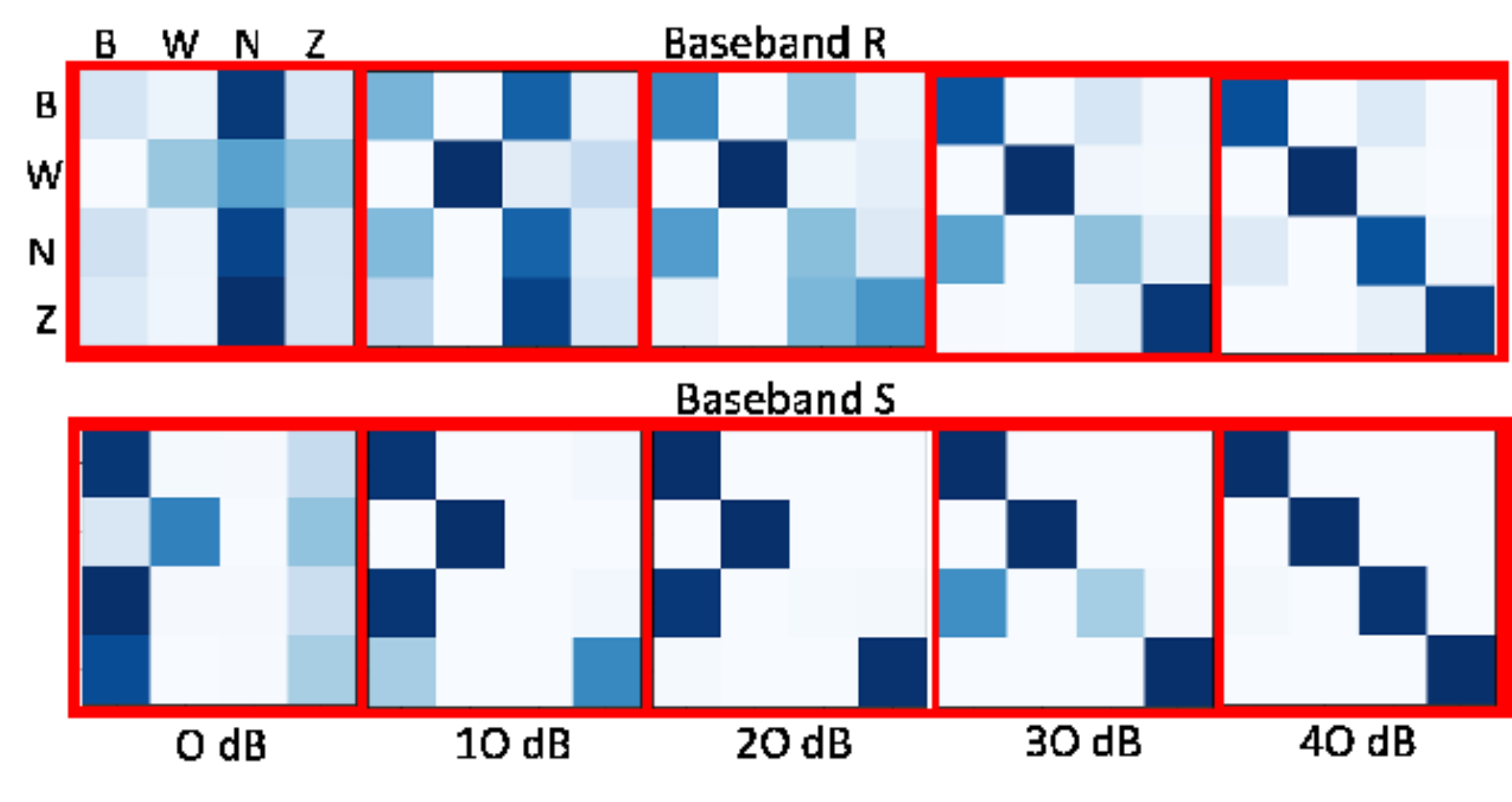}\vspace{-2mm}
\caption{{ Confusion matrices for ISMResNet11 trained on baseband random (top plots) and start bursts show (bottom) that at lower SNRs WiFi performs best, while BlueTooth and NRF are mixed up all the way to SNR of 40dB for random and 30dB for start bursts. ZigBee becomes recognizable in between those SNR values}}\vspace{-4mm}  \label{fig:f4a}
\end{center}
\end{figure}

\begin{figure}[t] %FIGURE 7
\begin{center}
%\begin{tabular}{c c}
%%\hspace{-5mm} \includegraphics [width=1.5in]{confmatrixAltResBT100Sw10.eps} & \includegraphics [width=1.5in]{confmatrixAltResBT100Sw50.eps} & \includegraphics [width=1.5in]{confmatrixAltResBT100Sw100.eps} %\vspace{-2mm}
%\hspace{-5mm} \includegraphics [width=1.5in]{confmatrixBTFC100Sw10.eps}  & \includegraphics [width=1.5in]{confmatrixBTFC100Sw100.eps} %\vspace{-2mm}
%\end{tabular}
\includegraphics [width=3.1in]{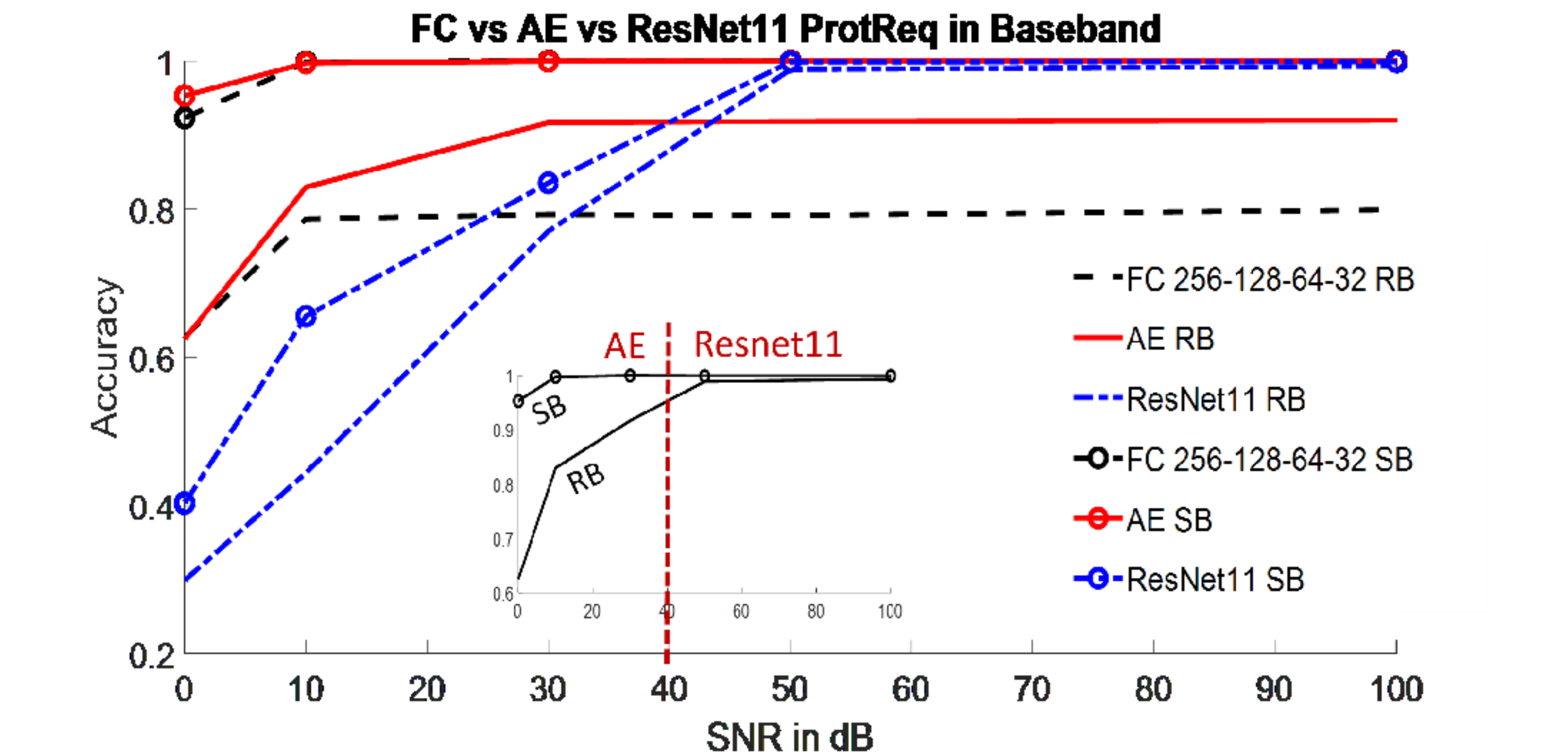}\vspace{-2mm}
\caption{{ Comparison of how different models perform across dataset SNRs; note the inserted graph that shows the accuracy for both start and random bursts if we switch from the AE to ISMResnet11 model for SNRs$>$40dB}}\vspace{-4mm}   \label{fig:f7}
\end{center}
\end{figure}

\begin{figure} %FIGURE 8
%\vspace{-4mm}
\begin{center}
\hspace{-5mm} \includegraphics [width=2.4in]{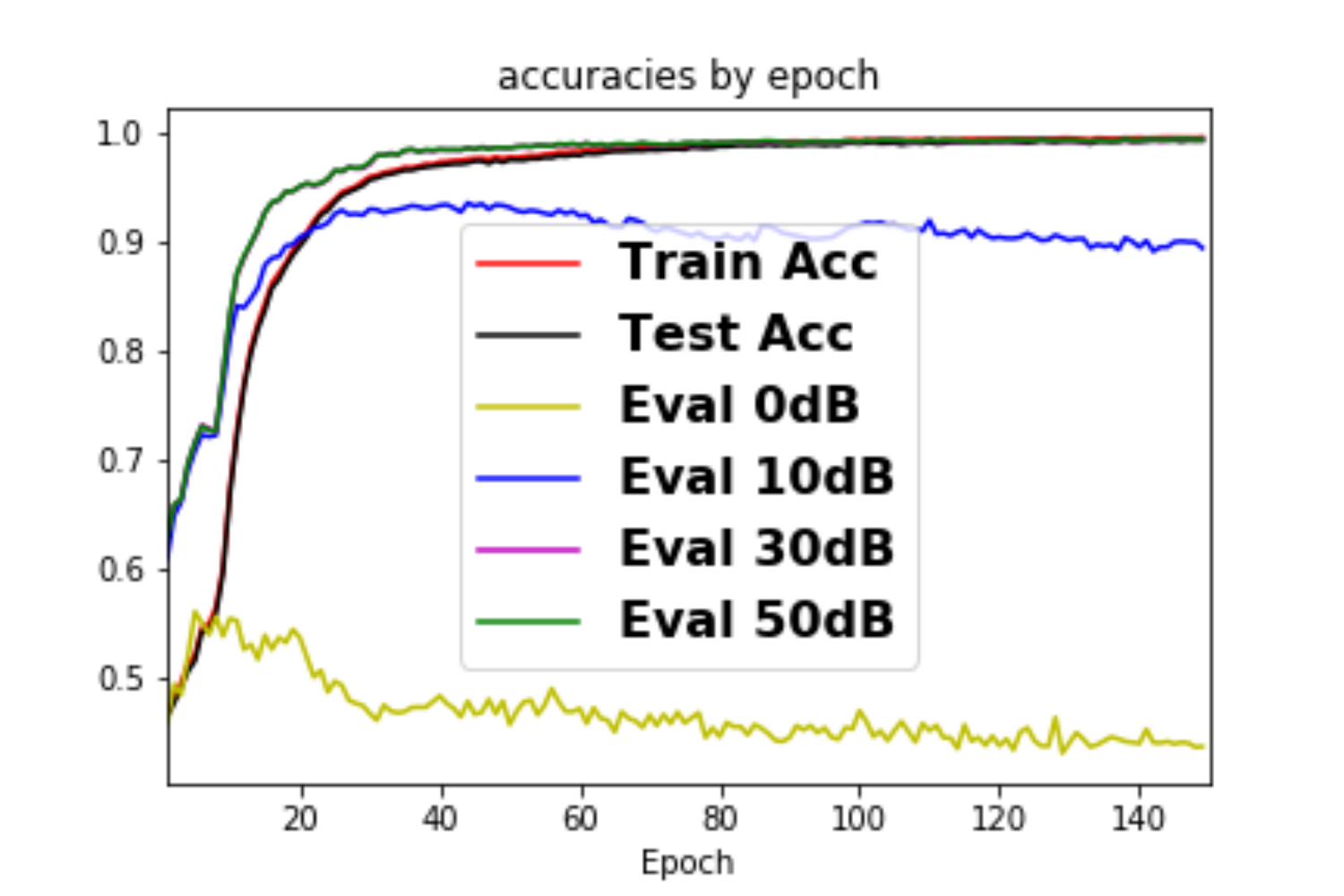} \vspace{-2mm}
\caption{The accuracy of SDAE-trained FC network over training epochs, evaluated on the training dataset of random bursts and randomized frequencies (plus several other datasets with same burst types but of lower SNRs)} \vspace{-5mm}
 \label{fig:f8} 
\end{center}
\end{figure}

\vspace{-1mm}
%%%%%%%%%%%%%%%%%%%%%%%%%%%%%%%%%%%%%%%%%%%%%%%%%%%%%%%%%%%%%%
Recall that data collection was performed in several recording campaigns, each lasting several seconds (some minutes) s.t. sampling the ISM signal produced contiguous sequence of I/Q samples, which we refer to as the recorded stream. We define the length $\ell$ of the burst to be sampled from the streams, and we use 2 different ways to sample the bursts: $\left.1\right)$ sample one burst of length $\ell$ from the start of each recorded stream of I/Q samples, $\left.2\right)$ sample bursts of length $\ell$ from random positions anywhere in recorded streams of I/Q samples.
%%%%%%%%%%%%%%%%%%%%%%%%%%%%%%%%%%%%%%%%%%%%%%%%%%%%%%%%%%%%%%
\begin{figure}[t] %FIGURE 4
\begin{center}
%\begin{tabular}{c c}
%%\hspace{-5mm} \includegraphics [width=1.5in]{confmatrixAltResBT100Sw10.eps} & \includegraphics [width=1.5in]{confmatrixAltResBT100Sw50.eps} & \includegraphics [width=1.5in]{confmatrixAltResBT100Sw100.eps} %\vspace{-2mm}
%\hspace{-5mm} \includegraphics [width=1.2in]{AltResConfusionMatrix0dBa.eps}  & \includegraphics [width=1.1in]{AltResConfusionMatrix50dB.eps} \vspace{-3mm}\\
%\hspace{-5mm} \includegraphics [width=1.2in]{AEConfusionmatrix0dBdrop13a.eps}  & \includegraphics [width=1.1in]{AEConfusionmatrixDrS50dB.eps} \vspace{-3mm}
%\end{tabular}
\vspace{-1mm}
\hspace{-5mm} \includegraphics [width=2.4in]{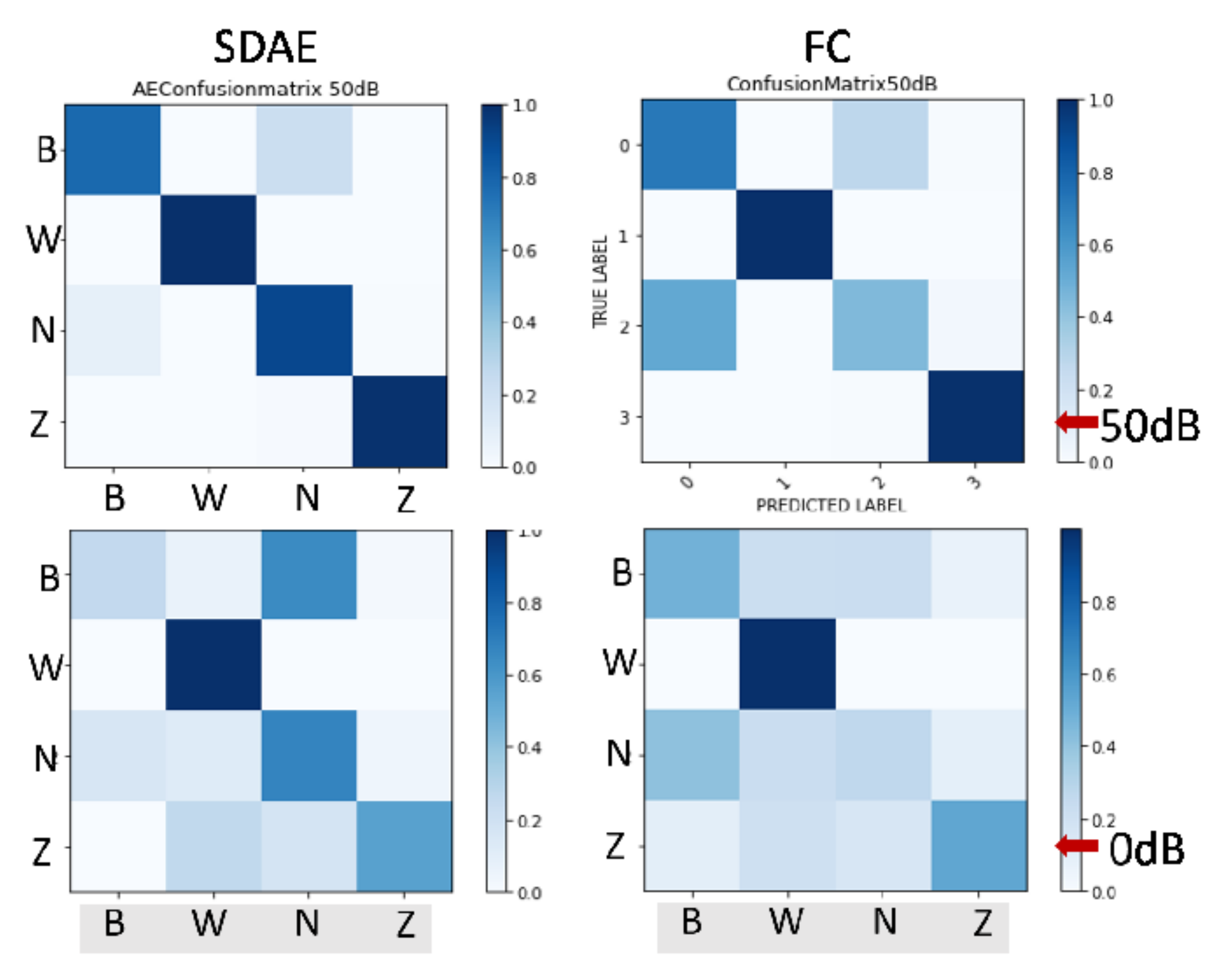}\vspace{-2mm}
\caption{{ Prediction results (0 and 50dB signals, bottom and top, respectively), based on random-burst-trained SDAE-196-96-20 (left) and random-burst-trained FC-256-128-64-32 (right), w/ classes coded as: 0-BT (B), 1-WiFi (W), 2-NRF (N), 3-ZBee (Z); Notice that the performance is similar at 0dB, while SDAE performs much better at 50 dB , as is seen in Fig.~\ref{fig:f7}, with the red and dashed black curves}} \vspace{-6mm}  \label{fig:f4}
\end{center}
\end{figure}
%\begin{itemize}
%\item sample one burst of length $\ell$ from the start of each recorded stream of I/Q samples,
%\item sample bursts of length $\ell$ from random positions anywhere in recorded streams of I/Q samples,
%\end{itemize}
{ We utilize both methods of burst sampling here. We used $\ell = 128$ across protocol recognition models, which is equivalent to $1.28 \mu$s of time and about 1 symbol for all featured protocols except WiFi where the burst covers multiple symbols. Start bursts are good to capture 
%transitional features of the signal that may reveal specific of the RF front end. For protocol classification, it's most 
important salient features such as sync pattern/preamble, and sometimes spread spectrum patterns. This is the reason that for each considered DL model better accuracy was achieved with start bursts only (observe curves marked with circles in Fig.~\ref{fig:f7}, and the bottom pane of Fig.~ \ref{fig:f4a}). Also, as featured in both  Fig.~ \ref{fig:f4} and Fig.~ \ref{fig:f4a}, confusion matrices for all models show how WiFi is the most resilient to low SNR, which is likely due to a very structured preamble. Bluetooth and NRF have the same GFSK modulation and same pulse shape; with the elimination of frequency band, there is not much left to distinguish them, hence at baseband  they are equally likely to be classified as one or another. Despite the similarity in modulation and pulse, we see that there are less dominant features that help overcome ambiguity for higher SNRs, and that the threshold SNR is lower for start bursts which indicates that preambles contain unique features. Contrary to the intuition from communication systems, ZigBee's Direct Sequence Spread Spectrum character does not make it resilient to low SNRs due to processing gain,  since the classifier is unaware of the spreading sequence. Fig.~ \ref{fig:f4} compares the reference FC network with the SDAE-based one in terms of the per-protocol robustness to noise when basebanded random bursts are used. The figure demonstrates that networks perform similarly at  0 dB, but the SDAE FC network performs much better at high SNRs, which is also reinforced by Fig.~ \ref{fig:f7}.  
  
Once the bursts are sampled, before performing the interleaving transform as described in Subsec.~\ref{subsubsec:data}, we add additional noise to emulate different receivers' SNR. This is how the evaluation datasets (0 - 50dB) in Fig.~\ref{fig:f8} were created. The noise added to evaluation datasets is different from the dusting noise used to train SDAE. An important observation is that both the results in Fig.~\ref{fig:f8} and the SDAE curves in Fig.~\ref{fig:f7} can be slightly modified using different dusting noise in each denoising AE utilized in SDAE. Different dusting may optimize the SDAE-trained FC to perform better at high SNRs, effectively reaching the performance of ISMResNet11, at the expense of the performance at 0 dB. If we want the classifier to exhibit maximum possible performance at all SNRs, and without the compactness constraint, we can keep the same dusting noise, and have a system that switches from SDAE-FC to ISMResNet11 at ~40dB, which would create a time-sharing performance bound represented by the insert in Fig.~\ref{fig:f7}.} 
%%%%%%%%%%%%%%%%%%%%%%%%%%%%%%%%%%%%%%%%%%%%%%%%%%%%%%%%%%%%%%
%\begin{figure}[t] %FIGURE 2
%\begin{center}
%\hspace{-5mm} \includegraphics [width=3.0in]{AltResLProtNoise100Acc4.eps}\vspace{-4mm}
%\caption{Shows how the accuracy changes with training steps for the ISMResNet11Residual network of 7 residual units per layer } \vspace{-2mm}  \label{fig:f2}
%\end{center}
%\end{figure}
%%%%%%%%%%%%%%%%%%%%%%%%%%%%%%%%%%%%%%%%%%%%%%%%%%%%%%%%%%%%%%%
%%%%%%%%%%%%%%%%%%%%%%%%%%%%%%%%%%%%%%%%%%%%%%%%%%%%%%%%%%%%%%%
 %\begin{figure}[t] %FIGURE 3
%\begin{center}
%\hspace{-5mm} \includegraphics [width=3.2in]{AltResLProtNoise100Loss4.eps}\vspace{-4mm}
%\caption{Shows how the loss function changes with training steps for the ISMResNet11 residual networks of seven residual units per layer.}   \label{fig:f3}
%\end{center}
%\end{figure}
%%%%%%%%%%%%%%%%%%%%%%%%%%%%%%%%%%%%%%%%%%%%%%%%%%%%%%%%%%%%%%%
%
%%%%%%%%%%%%%%%%%%%%%%%%%%%%%%%%%%%%%%%%%%%%%%%%%%%%%%%%%%%%%%
%%%%%%%%%%%%%%%%%%%%%%%%%%%%%%%%%%%%%%%%%%%%%%%%%%%%%%%%%%%%%%%
 %\begin{figure}[t] %FIGURE 5
%\begin{center}
%\hspace{-5mm} \includegraphics [width=3.0in]{FCProtacc.eps}  \vspace{-4mm}
%\caption{How the accuracy increases with training steps for the reference FC (256-128-64-32)}   \label{fig:f5}
%\end{center}
%\end{figure}
%%%%%%%%%%%%%%%%%%%%%%%%%%%%%%%%%%%%%%%%%%%%%%%%%%%%%%%%%%%%%%%
%
%%%%%%%%%%%%%%%%%%%%%%%%%%%%%%%%%%%%%%%%%%%%%%%%%%%%%%%%%%%%%%%
 %\begin{figure}[t] %FIGURE 6
%\begin{center}
%\hspace{-5mm} \includegraphics [width=3.0in]{FCProtloss.eps}  \vspace{-4mm}
%\caption{How the loss decreases with training steps for the reference FC network}   \label{fig:f6}
%\end{center}
%\end{figure}
%%%%%%%%%%%%%%%%%%%%%%%%%%%%%%%%%%%%%%%%%%%%%%%%%%%%%%%%%%%%%%%
%%%%%%%%%%%%%%%%%%%%%%%%%%%%%%%%%%%%%%%%%%%%%%%%%%%%%%%%%%%%%%
%%%%%%%%%%%%%%%%%%%%%%%%%%%%%%%%%%%%%%%%%%%%%%%%%%%%%%%%%%%%%%
As Fig.~\ref{fig:f1a} suggests, the idea followed in this paper is to use several bottleneck layers, obtained using successive DAEs, to obtain a robust and sparse low-dimensional input representation that could be easily separated by a classifier. 
%Fig.~\ref{fig:f1b} illustrates how each of the three bottleneck layers looked before the SDAE training, and what it looked like after the training. For the sake of visualization each datapoint presents only the three strongest eigenvalues of the given bottleneck vector. Note that despite this lossy but convenient quantization of bottleneck layers it could be observed that the transformed input space is more separable after the SDAE has been trained. Each point of this 3D representation of the bottleneck is colored according to the input class, hence the four colors.
%%%%%%%%%%%%%%%%%%%%%%%%%%%%%%%%%%%%%%%%%%%%
We designed the SDAE to use the loss $ \mathcal{L}$ defined in \ref{subsubsec:auto}. Some  parameter search is needed to identify the optimal $\rho$ for each particular transform. All 3 DAEs used the same $\rho$ and $\lambda,$ specific for interleaved transform. The DAEs inputs were dusted with independent noise of different variance. We call this the dusting noise, for the lack of a better term. 
If we applied more and more noise to DAE inputs, we would reach a point where all information is buried in noise, and no learning would occur. %Also, the more noise we used to dust the SDAE inputs, the less dropout regularization was needed. While both acted to randomize inputs and prevent overfitting, too much randomization would lower the information content entering a DAE.
After the SDAE was trained we retune the bottleneck layers stacked into the FC classifier. Adding some dropouts during retuning completely eliminates overfitting of the model. Training on start bursts resulted in the smallest footprint network (just two layers) with the best accuracy, even superior to ISMResNet11 for a range of SNRs (Fig.~\ref{fig:f7}). { Finally, there is a slight difference in the performance between randomized frequency channels and basebanded datapoints. Fig.~\ref{fig:f8} shows the accuracy during the training evaluated both on the training set and on several testing sets corrupted by noise, where all datasets had their frequency bands randomized. Close inspection of the final accuracy figures for all SNRs shows that they are slightly different than for baseband (Fig.~\ref{fig:f7}). Whether adjusting the levels of dusting noise for randomized frequencies would result in the accuracy-over-SNR curves that match baseband curves requires further investigation.}  Both reference and SDAE-based networks were trained on high SNR samples.
%%%%%%%%%%%%%%%%%%%%%%%%%%%%%%%%%%%%%%%%%%%%%%%%%%%%%%%%%%%%%%
%%%%%%%%%%%%%%%%%%%%%%%%%%%%%%%%%%%%%%%%%%%%%%%%%%%%%%%%%%%%%%
%%%%%%%%%%%%%%%%%%%%%%%%%%%%%%%%%%%%%%%%%%%%%%%%%%%%%%%%%%%%%%
 %\begin{figure}[t] %FIGURE 9
%\begin{center}
%\begin{tabular}{c c}
%%\hspace{-5mm} \includegraphics [width=1.5in]{confmatrixAltResBT100Sw10.eps} & \includegraphics [width=1.5in]{confmatrixAltResBT100Sw50.eps} & \includegraphics [width=1.5in]{confmatrixAltResBT100Sw100.eps} %\vspace{-2mm}
%\hspace{-5mm} \includegraphics [width=1.4in]{AEConfusionmatrixDrS0dB.eps}  & \includegraphics [width=1.4in]{AEConfusionmatrixDrS50dB.eps} \vspace{-3mm}
%\end{tabular}
%\caption{Prediction results for SDAE-196-96-20-D with starts of the burst} - WiFi again the most resilient, but improved prediction for  \vspace{-6mm}  \label{fig:f9}
%\end{center}
%\end{figure}
%%%%%%%%%%%%%%%%%%%%%%%%%%%%%%%%%%%%%%%%%%%%%%%%%%%%%%%%%%%%%%
\vspace{-2mm}
\section{Conclusion}%\vspace{-1mm}
We showed that compact FC networks trained to classify wireless protocols by using a hierarchy of 3 denoising autoencoders outperform reference FC networks trained in a typical way, i.e., with a stochastic gradient based optimization of the final FC model. %Each DAE corrupts the input with independent noise, where for the three FC layers trained by DAEs we used Gaussian noise of zero mean and three different standard deviations.
Not only is the complexity of such a FC network, measured in number of parameters and scalar multipliers, lower by at least $1/3$ than the reference FC, its accuracy also outperforms the larger FC for all tested SNR values (0 dB to 50dB). Depending on SNR, the new FC design also outperforms or closely approaches the performance of the referenced 3 times larger ResNet. We improved the classifier$'$s robustness to white noise while reducing its complexity.\vspace{-2mm}
%%Such SDAE-trained networks are suited for protocol inference performed by simple resource constrained devices based on noisy signal measurements. %Further research will extend this study to apply non-white noise in DAEs and to test the robustness and information loss of other signal transforms.
%%Compare with \cite{SparseDenoisingAEMigliori}
%
%%A Modulation Classification Method in Cognitive Radios System using Stacked Denoising Sparse Autoencoder
%
%
%%\subsection*{}
%
%%\cite{Merchant}
\bibliographystyle{IEEEtran}%
\bibliography{DLbib1}

\begin{thebibliography}{10}
\providecommand{\url}[1]{#1}
\csname url@samestyle\endcsname
\providecommand{\newblock}{\relax}
\providecommand{\bibinfo}[2]{#2}
\providecommand{\BIBentrySTDinterwordspacing}{\spaceskip=0pt\relax}
\providecommand{\BIBentryALTinterwordstretchfactor}{4}
\providecommand{\BIBentryALTinterwordspacing}{\spaceskip=\fontdimen2\font plus
\BIBentryALTinterwordstretchfactor\fontdimen3\font minus
  \fontdimen4\font\relax}
\providecommand{\BIBforeignlanguage}[2]{{%
\expandafter\ifx\csname l@#1\endcsname\relax
\typeout{** WARNING: IEEEtran.bst: No hyphenation pattern has been}%
\typeout{** loaded for the language `#1'. Using the pattern for}%
\typeout{** the default language instead.}%
\else
\language=\csname l@#1\endcsname
\fi
#2}}
\providecommand{\BIBdecl}{\relax}
\BIBdecl

\bibitem{80211n}
IEEE, ``{IEEE P802.11n™/D3.00 - Draft Amendment to STANDARD for Information
  Technology-Telecommunications and information exchange between Systems -
  Local and Metropolitan networks-Specific requirements-Part 11: Wireless LAN
  Medium Access Control (MAC) and Physical Layer (PHY). Amendment 4:
  Enhancements for Higher Throughput}.''

\bibitem{BT}
\BIBentryALTinterwordspacing
B.~S.~I. Group, ``{Bluetooth Specification Version 5.0},'' accessed on
  10/25/2018. [Online]. Available:
  \url{https://www.bluetooth.com/specifications/bluetooth-core-specification}
\BIBentrySTDinterwordspacing

\bibitem{ZBee}
N.~Salman, I.~Rasool, and A.~H. Kemp, ``{Overview of the IEEE 802.15.4
  standards family for Low Rate Wireless Personal Area Networks},'' in
  \emph{7th Int. Symp. on Wireless Communication Systems}, 2010.

\bibitem{NRF}
\BIBentryALTinterwordspacing
N.~Semiconductor, ``{nRF24L01 Single Chip 2.4GHz Transceiver},'' 2018, accessed
  on 10/25/2018. [Online]. Available:
  \url{https://www.nordicsemi.com/eng/content/download/2730/34105/file/nRF24L01_Product_Specification_v2_0.pdf}
\BIBentrySTDinterwordspacing

\bibitem{VincentSDAE}
{P. Vincent Pascal et al.}, ``{Stacked Denoising Autoencoders: Learning Useful
  Representations in a Deep Network with a Local Denoising Criterion},''
  \emph{The Jour. of Machine Learning Research}, vol.~11, pp. 3371--3408, Dec.
  2010.

\bibitem{SigMF}
\BIBentryALTinterwordspacing
SigMf, ``{The Signal Metadata Format Specification},'' accessed on 10/25/2018.
  [Online]. Available: \url{https://github.com/gnuradio/SigMF}
\BIBentrySTDinterwordspacing

\bibitem{PHYDeep}
T.~OShea and J.~Hoydis, ``{An introduction to deep learning for the physical
  layer},'' \emph{IEEE Transactions on Cognitive Communications and
  Networking}, 2017.

\bibitem{ConvDeepRF}
T.~J. OShea, J.~Corgan, and T.~C. Clancy, ``Convolutional radio modulation
  recognition networks,'' in \emph{Int. Conf. on Engineering Applications of
  Neural Networks}, 2016.

\bibitem{OTADeepRF}
{T. OShea et al.}, ``{Over the Air Deep Learning Based Radio Signal
  Classification},'' \emph{IEEE Journ. of Sel. Topics in Sig. Processing},
  2018.

\bibitem{cyclostat}
C.~M. Spooner, A.~N. Mody, J.~Chuang, and J.~Petersen, ``Modulation recognition
  using second-and higher-order cyclostationarity,'' in \emph{Int. Symp. on
  Dynamic Spectrum Access Networks (IEEE DySPAN)}, 2017.

\bibitem{GNUdataset}
T.~J. OShea and N.~West, ``Radio machine learning dataset generation with {GNU}
  radio,'' in \emph{GNU Radio Conference, vol. 1}, 2016.

\bibitem{Merchant}
{K. Merchant et al.}, ``Deep learning for rf device fingerprinting in cognitive
  communication networks,'' \emph{IEEE Journal of Selected Topics in Signal
  Processing}, 2018.

\bibitem{ResNet}
\BIBentryALTinterwordspacing
A.~Tamang, ``{ Another ResNet Tutorial (or not)},'' accessed on 10/25/2018.
  [Online]. Available:
  \url{https://medium.com/\@apiltamang/yet-another-resnet-tutorial-or-not-f6dd9515fcd7}
\BIBentrySTDinterwordspacing

\bibitem{KHeRes}
{He K. et al.}, ``Deep residual learning for image recognition,'' in
  \emph{Computer Vision and Pattern Recognition (CVPR)}, 2016.

\bibitem{SparseDenoisingAEMigliori}
{ B. Migliori et al.}, ``{Biologically inspired radio signal feature extraction
  with sparse denoising autoencoders},'' \emph{in ArXiv}, 2016.

\end{thebibliography}
\end{document}